\documentclass[floatfix,aps,prl,showpacs,amsmath,nofootinbib,preprintnumbers,twocolumn]{revtex4}

\usepackage{graphicx}

\newcommand{\figurewidth}{3.2in}

\def\half{{1\over 2}}

\def\ap{\alpha'}
\def\half{{1\over 2}}
\def\({\left(}
\def\){\right)}
\def\[{\left[}
\def\]{\right]}

\def\e{\begin{equation}}
\def\q{\end{equation}}
\def\m{\begin{eqnarray}}
\def\n{\end{eqnarray}}

\begin{document}

\title{Theoretic Limits on the Equation of State Parameter of Quintessence}

\author{Qing-Guo Huang}\email{huangqg@kias.re.kr}

\affiliation{School of physics, Korea Institute for Advanced Study,
207-43, Cheongryangri-Dong, Dongdaemun-Gu, Seoul 130-722, Korea}

%\date{\today}

\begin{abstract}

The value of scalar field coupled to gravity should be less than the
Planck scale in the consistent theory of quantum gravity. It
provides a theoretic constraint on the equation of state parameter
for the quintessence. In some cases our theoretic constraints are
more stringent than the constraints from the present experiments.

\end{abstract}


\maketitle

%%%%%%%%%%%%%%%%%%%%%%%%%%%%%%%%%%%%%%%%
%%%%%%%%%%%%%%%%%%%%%%%%%%%%%%%%%%%%%%%

The accelerating cosmic expansion is first inferred from the
observations of distant type Ia supernovae \cite{Riess:1998cb}. It
indicates unexpected gravitational physics attributed to the
dominating presence of a dark energy with negative pressure. Some
other independent observations, such the cosmic microwave background
radiation (CMBR) and Sloan Digital Sky Survey (SDSS), also strongly
favor dark energy as the dominant component in the present
mass-energy budget of the Universe.

A simply candidate for the dark energy is Einstein's famous
cosmological constant $\rho_\Lambda$. Nowadays it is still
consistent with all of the observations. See the recent analysis of
the experiments in \cite{Riess:2006fw,Spergel:2006hy}. The action
for the description of the Universe takes the form \e S={M_p^2\over
2}\int d^4x\sqrt{-g}R+\int d^4x\sqrt{-g}{\cal
L}(\phi_i,\psi_j,A_\mu,...), \q where $M_p$ is the reduced Planck
scale, $\phi_i$ is the scalar field, $\psi_j$ is the Fermionic
field, $A_\mu$ is the gauge field and so on. However anything that
contributes to the energy density of the vacuum acts just like a
cosmological constant. If we treat these quantum fields
independently, there is a zero point energy coupled to gravity.
Summing the zero point energies of all normal modes of some field of
mass $m$ up to a wave number cutoff $\Lambda \gg m$ yields a vacuum
energy density \e \langle\rho \rangle=\int_0^\Lambda {4\pi
k^2dk\over (2\pi)^3}\half\sqrt{k^2+m^2}\simeq {\Lambda^4\over
16\pi^2}.\q If we believe that the Planck scale is a natural cutoff
for the quantum field theories, $\langle \rho
\rangle={M_p^4/(16\pi^2)}$ is much greater than the observed value
of the energy density of dark energy $\rho_D=10^{-123}M_p^4$. The
energy scale for the local effective field theory related to the
cosmological constant is roughly $10^{-3}$ eV. The puzzle is why the
vacuum energy is so small after including all of these
contributions. Another problem is why the energy density of the dark
energy is comparable to the matter energy density now (cosmic
coincidence problem). For a classic review see
\cite{Weinberg:1988cp}, for a recent nice review see
\cite{Carroll:2000fy}, and for a recent discussion see
\cite{Polchinski:2006gy}.

The theory of quantum gravity is needed to solve the cosmological
constant problem. String theory appears to be a consistent and
well-defined theory of quantum gravity. In \cite{ArkaniHamed:2006dz}
Arkani-Hamed et al. suggest that the gravity and other quantum field
theory cannot be treated independently in quantum gravity. For
instance, a new intrinsic UV cutoff $\Lambda=gM_p$ for the U(1)
gauge theory with coupling $g$ coupled to gravity arises in
four-dimensional Minkowski spacetime. They also check this
conjecture in a few examples in string theory. The other concerning
on this conjecture is \cite{Huang:2007mf}. This conjecture is
generalized to the asymptotical de Sitter spacetime
\cite{Huang:2006hc}. The Hubble parameter $H$ plays the role as the
IR cutoff for the effective field theory. Requiring that the IR
cutoff be less than the UV cutoff leads to an upper bound on the
cosmological constant $\rho_\Lambda \leq g^2M_p^4$
\cite{Huang:2006hc}. This conjecture has a simple explanation in
string theory: the string length $\sqrt{\ap}$ should be shorter than
the size of the cosmic horizon. It can be easily checked in the
brane world scenario. See \cite{Huang:2006hc} in detail. If there is
a U(1) gauge theory with incredibly small coupling $g\sim 10^{-60}$
in our universe, we can understand why the cosmological constant is
so small. Similarly a conjecture for the $\lambda \phi^4$ theory is
proposed as $\Lambda=\lambda^{1/2}M_p$ in the Minkowski spacetime
and $\rho_\Lambda\leq \lambda M_p^4$ in the asymptotical de Sitter
space \cite{Huang:2007gk}. This conjecture implies that the value of
$\phi$ cannot be larger than the Planck scale $M_p$ and the chaotic
inflation cannot be achieved. However this conjecture is limited to
$\lambda \phi^4$ theory. We propose a general conjecture that the
description of the scalar field theory breaks down in the
over-Planckian field space in \cite{Huang:2007qz} where several
examples in string theory are discussed. For the other arguments in
string theory to support this conjecture see \cite{Ooguri:2006in}.

If the observed dark energy is really a small positive cosmological
constant the ultimate future of our universe will be eternal de
Sitter space. This would mean not that the future is totally empty
space, but that the would will have all the features of an isolated
finite thermal cavity with finite temperature and entropy
 $ S_{dS}={\hbox{Horizon Area}\over 4G}$ \cite{Gibbons:1977mu}. The
entropy reaches it maximum value and the second law forbids any
further interesting history. But on a sufficiently long time scale,
large fluctuations will occur. For de Sitter space the Poincare
recurrences generally occur on a time scale exponentially large in
the thermal entropy of the system $e^{S_{dS}}$ \cite{Goheer:2002vf}.
Recently the authors in \cite{ArkaniHamed:2007ky} propose that in
the de Sitter space the description of the local effective field
theory breaks down after a time scale $S_{dS}$ which is much shorter
than the recurrence time. Another trouble with the positive
cosmological constant is that it does not appear possible to define
precise observables, at least none that can be measured by an
observer in the spacetime \cite{Witten:2001kn}.

Another source for an appropriate dark energy component is a single
slow-rolling scalar field called quintessence \cite{Peebles:1987ek}.
In an expanding universe, a spatially homogeneous canonical scalar
field with potential $V(\phi)$ and minimal coupling to gravity obeys
\e\ddot \phi+3H\dot \phi+V'(\phi)=0, \q where the dot and prime
denote the derivative with respect to the cosmic time and
quintessence $\phi$ respectively. The energy density is
$\rho_Q=\half{\dot \phi}^2+V(\phi)$, and the pressure is $p_Q=\half
{\dot \phi}^2-V(\phi)$, implying an equation of state parameter \e
w\equiv{p_Q\over \rho_Q}={\half {\dot \phi}^2-V(\phi)\over \half
{\dot \phi}^2+V(\phi)}, \label{eosp}\q which generally varies with
time. The range for the equation of state parameter for the
quintessence is $w\in [-1,1]$. The Hubble parameter $H$ acts as a
friction term. When the friction term is large enough, the field is
slowly rolls down its potential and ${\dot \phi}^2\ll V(\phi)$. Now
$w\simeq -1$ and the quintessence acts like a cosmological constant.

This picture relies on an application of low-energy effective field
theory to the quintessence. So the variation of quintessence should
be less than $M_p$. In this paper, we will investigate the theoretic
constraint on the equation of state parameter for the quintessence
due to the sub-Planckian excursion in the field space.

For simplicity of calculations we assume spatial flatness which is
motivated by theoretical considerations, such as inflation, and
observations. Our results can be easily generalized to the case with
a spatial curvature. The Hubble parameter is given by \e
H^2={\rho_{crit}\over 3M_p^2}={\rho_Q+\rho_m\over 3M_p^2}, \q where
$\rho_m(z)=\rho_m^0 (1+z)^3$ is the energy density of the dust-like
matter in our universe and $\rho_m^0$ is its energy density at
present. Here we normalize $a_0=1$ and the scale factor is related
to the redshift $z$ by $a=(1+z)^{-1}$. Using eq. (\ref{eosp}), we
find a relationship between the potential of quintessence and its
kinetic energy \e V(\phi)={{\dot\phi}^2\over 2}{1-w\over 1+w}. \q
The energy density takes the form \e \rho_Q=\half{\dot
\phi}^2+V(\phi)={{\dot \phi}^2\over 1+w}. \label{rw}\q In the whole
paper, we assume, without loss of generality, $V'<0$, so that $\dot
\phi>0$. Thus eq. (\ref{rw}) reads \e \dot
\phi=\sqrt{(1+w)\rho_Q}.\label{dpw}\q Integrating eq. (\ref{dpw}),
we obtain \m
{|\Delta \phi(z)|\over M_p}&=&\int_{\phi(z)}^{\phi(0)} d\phi/M_p\nonumber \\
&=&\int_0^{z}\sqrt{3[1+w(z')]\Omega_Q(z')}{dz'\over 1+z'},
\label{vph}\n which should be less than 1. Here we use
$Hdt=-{dz\over 1+z}$. The density parameter for the quintessence is
\e \Omega_Q={\rho_Q\over \rho_{crit}}={\rho_Q\over \rho_Q+\rho_m}.
\q

The energy conservation implies \e \dot \rho_Q+3H(\rho_Q+p_Q)=0.
\label{emc}\q Combining with eq. (\ref{eosp}), we solve (\ref{emc})
as \e \rho_Q(z)=\rho_Q^0\exp\(\int_0^z3{1+w(z')\over 1+z'}dz'\),
\label{dq}\q where $\rho_Q^0$ is the present energy density of
quintessence. Using (\ref{dq}), we obtain \e {1\over
\Omega_Q(z)}=1+{\Omega_m^0\over
\Omega_Q^0}(1+z)^3\exp\(-\int_0^z3{1+w(z')\over 1+z'}dz'\).
\label{mq}\q For the case with a spatial curvature, we only need to
add another term $ -{\Omega_k^0\over
\Omega_Q^0}(1+z)^2\exp\(-\int_0^z 3{1+w(z')\over 1+z'}dz'\)$ on the
right hand side of eq. (\ref{mq}). Here we set $\Omega_k^0=0$.

Our strategy is using the condition $|\Delta \phi(z)|/M_p<1$ to
constraint the equation of state parameter $w(z)$ for the
quintessence. Unfortunately, present dynamical dark energy models in
the literatures do not suggest a universal or fundamental parametric
form for $w(z)$. For recent review see \cite{Szydlowski:2006ay}. We
will investigate several typical parameterizations of $w(z)$. There
are also strong degeneracies in the effect of $w(z)$ and $\Omega_m$
on the expansion history. According to the literatures, we
reasonably set $\Omega_Q^0=0.72$ and $\Omega_m^0=0.28$.

\begin{center}I. $w=w_0=$const \end{center}

There are a few models of quintessence that predict an equation of
state parameter that is constant, different from the cosmological
($w=-1$). In this case we consider the variation of the quintessence
from now to the last scattering ($z_{rec}=1089$). Requiring $|\Delta
\phi(z_{rec})|/M_p<1$ yields $w=w_0\leq -0.738$.

In a spatially flat universe, the combination of WMAP and the
Supernova Legacy Survey (SNLS) data leads to a significant
constraint on the equation of state parameter for the dark energy
$w=-0.967_{-0.072}^{+0.073}$ \cite{Spergel:2006hy}. The theoretic
limit on the equation of state parameter for the quintessence is
consistent with the experiments.

\begin{center}II. $w=w_0+w_1z$\end{center}

In this case the equation of state parameter is a linear function of
the redshift. This parametrization is studied in
\cite{Cooray:1999da}. It is a good parametrization at a low
redshift. But in this form, $w(z)$ diverges, making it unsuitable at
high redshift. As we know, the redshift of the SN sample is less
than 2. For the consistence we require $|\Delta \phi(z=2)|/M_p<1$.
The theoretic constraints are $-1\leq w_0\leq -0.164$ and
$-0.417\leq w_1\leq 0.854$. Here we also consider the requirement
$w\in[-1,1]$ for the quintessence. A more explicit result is showed
in Fig. 1.
\begin{figure}[ht]
\centerline{\includegraphics[width=\figurewidth]{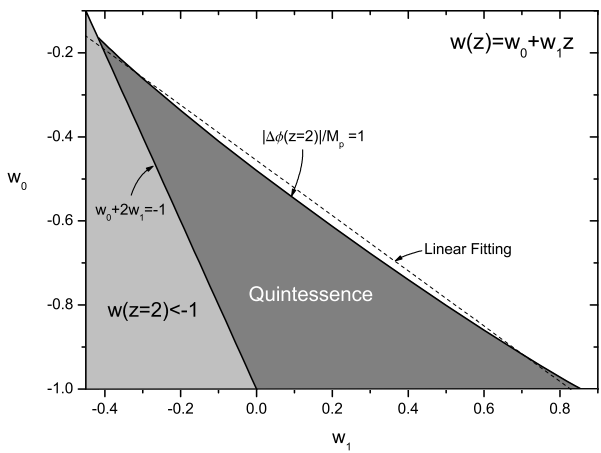}} \caption{The
gray patch is the prediction of the quintessence. The light gray
patch corresponds to $w<-1$. The line with $|\Delta \phi
(z=2)|/M_p=1$ is roughly a straight line which is linearly fitted as
$w_0+0.657w_1=-0.456$.}
\end{figure}

\begin{center}III. $w=w_0+w_1{z\over 1+z}$\end{center}

This parametric form is suggested in \cite{Chevallier:2000qy}. It
solves the divergence problem in case II and has been widely used in
the literatures. Requiring $|\Delta \phi(z_{rec})|/M_p<1$ yields
$-1\leq w_0\leq -0.434$ and $-0.564\leq w_1\leq 0.498$. See Fig. 2
for the explicit result.
\begin{figure}[ht]
\centerline{\includegraphics[width=\figurewidth]{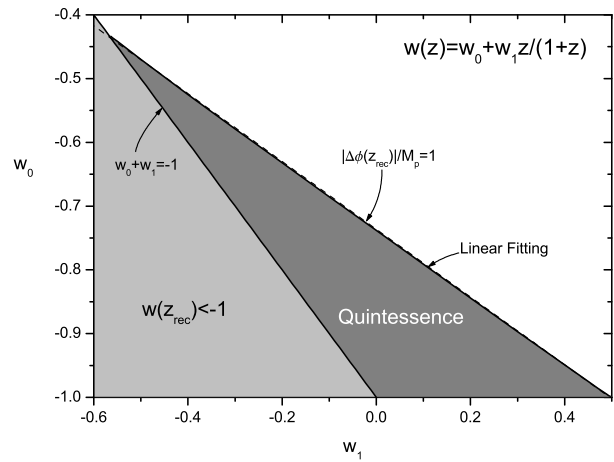}} \caption{The
gray patch is the prediction of the quintessence. The light gray
patch corresponds to $w<-1$. The line with $|\Delta \phi
(z=2)|/M_p=1$ is roughly a straight line which is linearly fitted as
$w_0+0.532w_1=-0.736$.}
\end{figure}

The fitting results for the combination of WMAP3 + SN182 + SDSS +
2dFGRS are $w_0=-1.149_{-0.120}^{+0.543}$ and
$w_1=1.017_{-2.095}^{+0.146}$ at $2\sigma$ level \cite{Zhao:2006qg}.
The theoretic constraint is consistent with the experiment as well.
Taking a closer look, we find the theoretic limit is much more
stringent than the present experiments.

%\begin{center}IV. $w=w_0+w_1\ln(1+z)$\end{center}

The authors in \cite{Caldwell:2005tm} propose that the equation of
state parameter has a lower bound, such as $1+w\geq 0.004$ or $0.01$
for two different cases, if the value of the quintessence field is
prohibited from attaining values exceeding the Planck scale. Their
results contradict with our intuition. The inconsistence in
\cite{Caldwell:2005tm} is that they require $E\equiv|V/V'|<M_p$.
Recall what we do for inflation. Similarly we define a slow-roll
parameter $\epsilon={M_p^2\over 2}(V'/V)^2=M_p^2/(2E^2)>1$, which
implies that the quintessence fast rolls down its potential and the
accelerating expansion cannot be achieved when this quintessence is
the dominant component in the Universe. In fact the absolute value
of the scalar field does not make sense because the scalar field can
be shifted to be arbitrary value. The reasonable quantity is the
value of the scalar field relative to its value in the closed-by
stable/metastable vacuum. So the requirement that the variation of
the quintessence be less than $M_p$ is more reasonable. The
variation of quintessence can be arbitrarily small if its potential
is flat enough, and the quintessence just acts like the cosmological
constant. The only possible lower bound on $w$ is
$1+w>H_0^2/M_p^2\sim 10^{-120}$; otherwise, the current inflationary
epoch is eternal \cite{ArkaniHamed:2007ky}.

To summarize, we investigate the theoretic limits on the equation of
state parameter for quintessence through considering that the
variation of the canonical quintessence field minimally coupled to
gravity is less than the Planck scale. This requirement may arise in
the consistent theory of quantum gravity. In this sense our results
can be taken as the prediction of quantum gravity. Our theoretic
constraints are more stringent than present experiments in some
cases. We hope the future observations can confirm the quintessence
model or rule it out.

However there is also a new naturalness problem in the quintessence
model. In order for the quintessence to be slowly rolling today, the
effective mass of quintessence should be smaller than the present
Hubble parameter, namely $m_\phi=\sqrt{|V''(\phi_0)|}\leq H_0\sim
10^{-33}$ eV. This is an incredibly low energy scale compared to the
energy scales in particle physics. If so, there must be an unknown
symmetry to protect such a tiny mass of quintessence.

In ten dimensions string theory has no free parameters, but once we
compactify, each nonsupersymmetric vacuum will have a different
effective cosmological constant \cite{Kachru:2003aw,Krause:2007gj}.
In the reliable set-up in string theory \cite{Kachru:2003aw}, the
vacuum with a positive cosmological constant is metastable. The
trouble with the eternal de Sitter space we discussed previously
seems to be solved. It hints that we should embed the pure gravity
with a positive cosmological constant into a bigger theory. Since
the cosmological constant is not a dynamical quantity, maybe
anthropic principle \cite{Weinberg:1987dv} is needed.

Nowadays the dark energy with $w<-1$, called phantom
\cite{Caldwell:1999ew}, has not been ruled out by the experiments.
If phantom is favored by the future observations, it is a bigger
puzzle for the fundamental physics, because the Null energy
condition (NEC) is violated, which implies that energy flows faster
than the speed of light \cite{Hawking:1973uf}. The causality is
absent as well. A simple realization of phantom is the scalar field
with a wrong sign kinetic term \cite{Singh:2003vx}. We don't know
how to quantize the phantom field at all and the field theory of
phantom is ill-defined. To constrain the equation of state parameter
for phantom is out of the question. On the other hand, NEC is a
crucial assumption in proving the positivity of the ADM mass in
asymptotically flat space \cite{Schon:1981vd}. The positive energy
theorem implies a stable vacuum for gravity and will play a crucial
role in quantum gravity.

The cosmological constant problem is still the biggest puzzle in the
fundamental physics. We are still far away from the correct answer
to it.

\noindent {\bf Acknowledgments.} We would like to thank F.~L.~Lin
and P.~J.~Yi for useful discussions.

%%%%%%%%%%%%%%%%%%%%%%%%%%%%%%%%%%%%%%%%
%%%%%%%%%%%%%%%%%%%%%%%%%%%%%%%%%%%%%%%%

%%%%%%%%%%%%%%%%%%%%%%%%%%%%%%%%%%%%%%%%
%%%%%%%%%%%%%%%%%%%%%%%%%%%%%%%%%%%%%%%%

%%%%%%%%%%%%%%%%%%%%%%%%%%%%%%%%%%%%%%%%
%%%%%%%%%%%%%%%%%%%%%%%%%%%%%%%%%%%%%%%%

\begin{thebibliography}{99}
\frenchspacing
%%%%%%%%%%%%%%%%%%%%%%%%%%%%%%%%%%%%%%%%
%%%%%%%%%%%%%%%%%%%%%%%%%%%%%%%%%%%%%%%%

%\cite{Riess:1998cb}
\bibitem{Riess:1998cb}
  A.~G.~Riess {\it et al.}  [Supernova Search Team Collaboration],
  %``Observational Evidence from Supernovae for an Accelerating Universe and a
  %Cosmological Constant,''
  Astron.\ J.\  {\bf 116}, 1009 (1998)
  [arXiv:astro-ph/9805201];
  %%CITATION = ANJOA,116,1009;%%
%\cite{Perlmutter:1998np}
%\bibitem{Perlmutter:1998np}
  S.~Perlmutter {\it et al.}  [Supernova Cosmology Project Collaboration],
  %``Measurements of Omega and Lambda from 42 High-Redshift Supernovae,''
  Astrophys.\ J.\  {\bf 517}, 565 (1999)
  [arXiv:astro-ph/9812133].
  %%CITATION = ASJOA,517,565;%%

%\cite{Riess:2006fw}
\bibitem{Riess:2006fw}
  A.~G.~Riess {\it et al.},
  %``New Hubble Space Telescope Discoveries of Type Ia Supernovae at $z > 1$:
  %Narrowing Constraints on the Early Behavior of Dark Energy,''
  arXiv:astro-ph/0611572.
  %%CITATION = ASTRO-PH/0611572;%%

%\cite{Spergel:2006hy}
\bibitem{Spergel:2006hy}
  D.~N.~Spergel {\it et al.}  [WMAP Collaboration],
  %``Wilkinson Microwave Anisotropy Probe (WMAP) three year results:
  %Implications for cosmology,''
  Astrophys.\ J.\ Suppl.\  {\bf 170}, 377 (2007)
  [arXiv:astro-ph/0603449].
  %%CITATION = APJSA,170,377;%%

%\cite{Weinberg:1988cp}
\bibitem{Weinberg:1988cp}
  S.~Weinberg,
  %``The cosmological constant problem,''
  Rev.\ Mod.\ Phys.\  {\bf 61}, 1 (1989).
  %%CITATION = RMPHA,61,1;%%

%\cite{Carroll:2000fy}
\bibitem{Carroll:2000fy}
  S.~M.~Carroll,
  %``The cosmological constant,''
  Living Rev.\ Rel.\  {\bf 4}, 1 (2001)
  [arXiv:astro-ph/0004075].
  %%CITATION = 00222,4,1;%%

%\cite{Polchinski:2006gy}
\bibitem{Polchinski:2006gy}
  J.~Polchinski,
  %``The cosmological constant and the string landscape,''
  arXiv:hep-th/0603249.
  %%CITATION = HEP-TH/0603249;%%

%\cite{ArkaniHamed:2006dz}
\bibitem{ArkaniHamed:2006dz}
  N.~Arkani-Hamed, L.~Motl, A.~Nicolis and C.~Vafa,
  %``The string landscape, black holes and gravity as the weakest force,''
  JHEP {\bf 0706}, 060 (2007)
  [arXiv:hep-th/0601001].
  %%CITATION = JHEPA,0706,060;%%

\bibitem{Huang:2007mf}
  Q.~G.~Huang,
  %``Gravitational correction and weak gravity conjecture,''
  JHEP {\bf 0703}, 053 (2007)
  [arXiv:hep-th/0703039];
  %%CITATION = JHEPA,0703,053;%%
%\cite{Banks:2006mm}
%\bibitem{Banks:2006mm}
  T.~Banks, M.~Johnson and A.~Shomer,
  %``A note on gauge theories coupled to gravity,''
  JHEP {\bf 0609}, 049 (2006)
  [arXiv:hep-th/0606277].
  %%CITATION = JHEPA,0609,049;%%
%\cite{Huang:2007mf}


%\cite{Huang:2006hc}
\bibitem{Huang:2006hc}
  Q.~G.~Huang, M.~Li and W.~Song,
  %``Bound on the U(1) gauge coupling in the asymptotically dS and AdS
  %background,''
  JHEP {\bf 0610}, 059 (2006)
  [arXiv:hep-th/0603127].
  %%CITATION = JHEPA,0610,059;%%

%\cite{Huang:2007gk}
\bibitem{Huang:2007gk}
  Q.~G.~Huang,
  %``Weak gravity conjecture constraints on inflation,''
  JHEP {\bf 0705}, 096 (2007)
  [arXiv:hep-th/0703071].
  %%CITATION = JHEPA,0705,096;%%

%\cite{Huang:2007qz}
\bibitem{Huang:2007qz}
  Q.~G.~Huang,
  %``Constraints on the spectral index for the inflation models in string
  %landscape,''
  arXiv:0706.2215 [hep-th].
  %%CITATION = ARXIV:0706.2215;%%

%\cite{Ooguri:2006in}
\bibitem{Ooguri:2006in}
  H.~Ooguri and C.~Vafa,
  %``On the geometry of the string landscape and the swampland,''
  Nucl.\ Phys.\  B {\bf 766}, 21 (2007)
  [arXiv:hep-th/0605264].
  %%CITATION = NUPHA,B766,21;%%

%\cite{Gibbons:1977mu}
\bibitem{Gibbons:1977mu}
  G.~W.~Gibbons and S.~W.~Hawking,
  %``Cosmological Event Horizons, Thermodynamics, And Particle Creation,''
  Phys.\ Rev.\  D {\bf 15}, 2738 (1977).
  %%CITATION = PHRVA,D15,2738;%%

%\cite{Goheer:2002vf}
\bibitem{Goheer:2002vf}
  N.~Goheer, M.~Kleban and L.~Susskind,
  %``The trouble with de Sitter space,''
  JHEP {\bf 0307}, 056 (2003)
  [arXiv:hep-th/0212209];
  %%CITATION = JHEPA,0307,056;%%
%\cite{Susskind:2003kw}
%\bibitem{Susskind:2003kw}
  L.~Susskind,
  %``The anthropic landscape of string theory,''
  arXiv:hep-th/0302219.
  %%CITATION = HEP-TH/0302219;%%

%\cite{ArkaniHamed:2007ky}
\bibitem{ArkaniHamed:2007ky}
  N.~Arkani-Hamed, S.~Dubovsky, A.~Nicolis, E.~Trincherini and G.~Villadoro,
  %``A Measure of de Sitter Entropy and Eternal Inflation,''
  JHEP {\bf 0705}, 055 (2007)
  [arXiv:0704.1814 [hep-th]].
  %%CITATION = JHEPA,0705,055;%%

%\cite{Witten:2001kn}
\bibitem{Witten:2001kn}
  E.~Witten,
  %``Quantum gravity in de Sitter space,''
  arXiv:hep-th/0106109.
  %%CITATION = HEP-TH/0106109;%%


%\cite{Peebles:1987ek}
\bibitem{Peebles:1987ek}
  P.~J.~E.~Peebles and B.~Ratra,
  %``Cosmology with a Time Variable Cosmological Constant,''
  Astrophys.\ J.\  {\bf 325}, L17 (1988);
  %%CITATION = ASJOA,325,L17;%%
%\cite{Wetterich:1994bg}
%\bibitem{Wetterich:1994bg}
  C.~Wetterich,
  %``The Cosmon model for an asymptotically vanishing time dependent
  %cosmological 'constant',''
  Astron.\ Astrophys.\  {\bf 301}, 321 (1995)
  [arXiv:hep-th/9408025];
  %%CITATION = AAEJA,301,321;%%
%\cite{Caldwell:1997ii}
%\bibitem{Caldwell:1997ii}
  R.~R.~Caldwell, R.~Dave and P.~J.~Steinhardt,
  %``Cosmological Imprint of an Energy Component with General
  %Equation-of-State,''
  Phys.\ Rev.\ Lett.\  {\bf 80}, 1582 (1998)
  [arXiv:astro-ph/9708069];
  %%CITATION = PRLTA,80,1582;%%
%\cite{Huey:1998se}
%\bibitem{Huey:1998se}
  G.~Huey, L.~M.~Wang, R.~Dave, R.~R.~Caldwell and P.~J.~Steinhardt,
  %``Resolving the Cosmological Missing Energy Problem,''
  Phys.\ Rev.\  D {\bf 59}, 063005 (1999)
  [arXiv:astro-ph/9804285];
  %%CITATION = PHRVA,D59,063005;%%
%\cite{Peebles:2002gy}
%\bibitem{Peebles:2002gy}
  P.~J.~E.~Peebles and B.~Ratra,
  %``The cosmological constant and dark energy,''
  Rev.\ Mod.\ Phys.\  {\bf 75}, 559 (2003)
  [arXiv:astro-ph/0207347].
  %%CITATION = RMPHA,75,559;%%

%\cite{Szydlowski:2006ay}
\bibitem{Szydlowski:2006ay}
  M.~Szydlowski, A.~Kurek and A.~Krawiec,
  %``Top ten accelerating cosmological models,''
  Phys.\ Lett.\  B {\bf 642}, 171 (2006)
  [arXiv:astro-ph/0604327];
  %%CITATION = PHLTA,B642,171;%%
%\cite{Copeland:2006wr}
%\bibitem{Copeland:2006wr}
  E.~J.~Copeland, M.~Sami and S.~Tsujikawa,
  %``Dynamics of dark energy,''
  Int.\ J.\ Mod.\ Phys.\  D {\bf 15}, 1753 (2006)
  [arXiv:hep-th/0603057];
  %%CITATION = IMPAE,D15,1753;%%
%\cite{Wei:2006ut}
%\bibitem{Wei:2006ut}
  H.~Wei and S.~N.~Zhang,
  %``Observational $H(z)$ Data and Cosmological Models,''
  Phys.\ Lett.\  B {\bf 644}, 7 (2007)
  [arXiv:astro-ph/0609597].
  %%CITATION = PHLTA,B644,7;%%

%\cite{Cooray:1999da}
\bibitem{Cooray:1999da}
  A.~R.~Cooray and D.~Huterer,
  %``Gravitational Lensing as a Probe of Quintessence,''
  Astrophys.\ J.\  {\bf 513}, L95 (1999)
  [arXiv:astro-ph/9901097];
  %%CITATION = ASJOA,513,L95;%%
%\cite{Di Pietro:2002cz}
%\bibitem{DiPietro:2002cz}
  E.~Di Pietro and J.~F.~Claeskens,
  %``Quintessence models faced with future supernovae data,''
  Mon.\ Not.\ Roy.\ Astron.\ Soc.\  {\bf 341}, 1299 (2003)
  [arXiv:astro-ph/0207332].
  %%CITATION = MNRAA,341,1299;%%


%\cite{Chevallier:2000qy}
\bibitem{Chevallier:2000qy}
  M.~Chevallier and D.~Polarski,
  %``Accelerating universes with scaling dark matter,''
  Int.\ J.\ Mod.\ Phys.\  D {\bf 10}, 213 (2001)
  [arXiv:gr-qc/0009008];
  %%CITATION = IMPAE,D10,213;%%
%\cite{Linder:2002et}
%\bibitem{Linder:2002et}
  E.~V.~Linder,
  %``Exploring the expansion history of the universe,''
  Phys.\ Rev.\ Lett.\  {\bf 90}, 091301 (2003)
  [arXiv:astro-ph/0208512].
  %%CITATION = PRLTA,90,091301;%%


%\cite{Zhao:2006qg}
\bibitem{Zhao:2006qg}
  G.~B.~Zhao, J.~Q.~Xia, H.~Li, C.~Tao, J.~M.~Virey, Z.~H.~Zhu and X.~Zhang,
  %``Probing for dynamics of dark energy and curvature of universe with latest
  %cosmological observations,''
  Phys.\ Lett.\  B {\bf 648}, 8 (2007)
  [arXiv:astro-ph/0612728].
  %%CITATION = PHLTA,B648,8;%%


%\cite{Caldwell:2005tm}
\bibitem{Caldwell:2005tm}
  R.~R.~Caldwell and E.~V.~Linder,
  %``The limits of quintessence,''
  Phys.\ Rev.\ Lett.\  {\bf 95}, 141301 (2005)
  [arXiv:astro-ph/0505494].
  %%CITATION = PRLTA,95,141301;%%

%\cite{Kachru:2003aw}
\bibitem{Kachru:2003aw}
  S.~Kachru, R.~Kallosh, A.~Linde and S.~P.~Trivedi,
  %``De Sitter vacua in string theory,''
  Phys.\ Rev.\  D {\bf 68}, 046005 (2003)
  [arXiv:hep-th/0301240].
  %%CITATION = PHRVA,D68,046005;%%

%\cite{Krause:2007gj}
\bibitem{Krause:2007gj}
  A.~Krause,
  %``Supersymmetry breaking with zero vacuum energy in M-theory flux
  %compactifications,''
  Phys.\ Rev.\ Lett.\  {\bf 98}, 241601 (2007)
  [arXiv:hep-th/0701009].
  %%CITATION = PRLTA,98,241601;%%

%\cite{Weinberg:1987dv}
\bibitem{Weinberg:1987dv}
  S.~Weinberg,
  %``Anthropic Bound on the Cosmological Constant,''
  Phys.\ Rev.\ Lett.\  {\bf 59}, 2607 (1987);
  %%CITATION = PRLTA,59,2607;%%
%\cite{Bousso:2007kq}
%\bibitem{Bousso:2007kq}
  R.~Bousso, R.~Harnik, G.~D.~Kribs and G.~Perez,
  %``Predicting the Cosmological Constant from the Causal Entropic Principle,''
  arXiv:hep-th/0702115.
  %%CITATION = HEP-TH/0702115;%%

%\cite{Caldwell:1999ew}
\bibitem{Caldwell:1999ew}
  R.~R.~Caldwell,
  %``A Phantom Menace?,''
  Phys.\ Lett.\  B {\bf 545}, 23 (2002)
  [arXiv:astro-ph/9908168].
  %%CITATION = PHLTA,B545,23;%%

%\cite{Hawking:1973uf}
\bibitem{Hawking:1973uf}
  S.~W.~Hawking and G.~F.~R.~Ellis,
  ``The Large scale structure of space-time,''
%\href{http://www.slac.stanford.edu/spires/find/hep/www?irn=6991262}{SPIRES entry}
{\it  Cambridge University Press, Cambridge, 1973};
%\cite{Wald:1984rg}
%\bibitem{Wald:1984rg}
  R.~M.~Wald,
  ``General Relativity,''
%\href{http://www.slac.stanford.edu/spires/find/hep/www?irn=1334239}{SPIRES entry}
{\it  Chicago, Usa: Univ. Pr. ( 1984) 491p}.

%\cite{Singh:2003vx}
\bibitem{Singh:2003vx}
  P.~Singh, M.~Sami and N.~Dadhich,
  %``Cosmological dynamics of phantom field,''
  Phys.\ Rev.\  D {\bf 68}, 023522 (2003)
  [arXiv:hep-th/0305110];
  %%CITATION = PHRVA,D68,023522;%%
%\cite{Nojiri:2005pu}
%\bibitem{Nojiri:2005pu}
  S.~Nojiri and S.~D.~Odintsov,
  %``Unifying phantom inflation with late-time acceleration: Scalar
  %phantom-non-phantom transition model and generalized holographic dark
  %energy,''
  Gen.\ Rel.\ Grav.\  {\bf 38}, 1285 (2006)
  [arXiv:hep-th/0506212];
  %%CITATION = GRGVA,38,1285;%%
%\cite{Capozziello:2005tf}
%\bibitem{Capozziello:2005tf}
  S.~Capozziello, S.~Nojiri and S.~D.~Odintsov,
  %``Unified phantom cosmology: Inflation, dark energy and dark matter under
  %the same standard,''
  Phys.\ Lett.\  B {\bf 632}, 597 (2006)
  [arXiv:hep-th/0507182].
  %%CITATION = PHLTA,B632,597;%%
%\cite{Hrycyna:2007gd}
%\bibitem{Hrycyna:2007gd}
  O.~Hrycyna and M.~Szydlowski,
  %``Extended Quintessence with non-minimally coupled phantom scalar field,''
  arXiv:0707.4471 [hep-th].
  %%CITATION = ARXIV:0707.4471;%%


%\cite{Schon:1981vd}
\bibitem{Schon:1981vd}
  R.~Schon and S.~T.~Yau,
  %``Proof Of The Positive Mass Theorem. 2,''
  Commun.\ Math.\ Phys.\  {\bf 79}, 231 (1981);
  %%CITATION = CMPHA,79,231;%%
%\cite{Witten:1981mf}
%\bibitem{Witten:1981mf}
  E.~Witten,
  %``A Simple Proof Of The Positive Energy Theorem,''
  Commun.\ Math.\ Phys.\  {\bf 80}, 381 (1981).
  %%CITATION = CMPHA,80,381;%%


%%%%%%%%%%%%%%%%%%%%%%%%%%%%%%%%%%%%%%%%
%%%%%%%%%%%%%%%%%%%%%%%%%%%%%%%%%%%%%%%%
\end{thebibliography}
\end{document}